\documentclass[11pt, reqno]{article}
\pdfoutput=1

\usepackage{graphicx}
\usepackage{jheppub}
\usepackage{amssymb}
\usepackage{amsmath}
\usepackage[usenames,dvipsnames]{xcolor}
\usepackage{epsfig}
\usepackage{dcolumn}
\usepackage{tikz}
\usetikzlibrary{shapes.geometric, arrows}
\usepackage{upgreek}
\usepackage{setspace}
\usepackage{lmodern}
\usepackage{enumitem}
\usepackage{array,multirow,bigdelim,arydshln}
\usepackage{appendix}
\usepackage{xparse}
\usepackage[utf8]{inputenc}
\usepackage{hyperref}
\usepackage[dvipsnames]{xcolor}
\usepackage{blkarray}
\hypersetup{
	colorlinks,
	urlcolor=Maroon,
	linkcolor=Maroon,
	citecolor=Maroon
}

\usepackage{amsthm}
\newtheorem{thm}{Theorem}[section]
\newtheorem{prop}[thm]{Proposition}

\theoremstyle{definition}
\newtheorem{example}[thm]{Example}

\usepackage{mathtools}

\usepackage{float}
\restylefloat{table}

\NewDocumentCommand{\binomial}{omm}
 {%
  \genfrac(){0pt}{}{#2}{#3}%
  \IfValueT{#1}{_{\!#1}}%
 }
\NewDocumentCommand{\eulerian}{omm}
 {%
  \genfrac<>{0pt}{}{#2}{#3}%
  \IfValueT{#1}{_{\!#1}}%
 }

\def \s {\sigma}

\usepackage{underscore}

\usepackage{latexsym}
\usepackage{tikz}

\title{Splitting CEGM Amplitudes}

\author[a]{Bruno Gim\'enez Umbert}\emailAdd{b.gimenez-umbert@soton.ac.uk}
\author[b]{and Bernd Sturmfels}\emailAdd{bernd@mis.mpg.de}
\affiliation[a]{Physics \& Astronomy and Mathematical Sciences, University of Southampton, UK}
\affiliation[b]{Max Planck Institute for Mathematics in the Sciences, Leipzig, Germany}

\abstract{The CEGM formalism offers a general framework for
scattering amplitudes, which rests on Grassmannians, moduli spaces and tropical geometry.
The physical implications of this generalization are still to be understood.
  Conventional wisdom says that key features of scattering amplitudes,
  like factorization at their poles into lower-point amplitudes, are associated to their singularities.
  The factorization behavior of CEGM amplitudes at their poles is interesting but complicated.
    Recent developments have revealed important properties of standard particle and string scattering 
    amplitudes from factorizations,
    known as splits, that happen away from poles.
In this paper we introduce a kinematic subspace on which the CEGM amplitude  splits into
very simple rational functions. These functions, called simplex amplitudes,
 arise from stringy integrals for the multivariate beta function, and also from restricting
the biadjoint scalar amplitude in quantum field theory to
  certain kinematic loci. Using split kinematics we also discover a specific class of zeros of the CEGM amplitude.
Our construction rests on viewing positive moduli space as a product of simplices,
and it suggests a novel approach for deriving scattering amplitudes
from tropical  determinantal varieties. }

\begin{document}
\maketitle
\addtocontents{toc}{\protect\setcounter{tocdepth}{1}}
\def \tr {\nonumber\\}
\def \nn {\nonumber}
\def \la {|}
\def \ra {|}
\def \dd {\Theta}
\def\hset{\texttt{h}}
\def\gset{\texttt{g}}
\def\sset{\texttt{s}}
\def \be {\begin{equation}}
\def \ee {\end{equation}}
\def \ba {\begin{eqnarray}}
\def \ea {\end{eqnarray}}
\def \k {\kappa}
\def \h {\hbar}
\def \r {\rho}
\def \l {\lambda}
\def \be {\begin{equation}}
\def \en {\end{equation}}
\def \bes {\begin{eqnarray}}
\def \ens {\end{eqnarray}}
\def \red {\color{Maroon}}
\def \pt {{\rm PT}}
\def \s {\textsf{s}}
\def \t {\textsf{t}}
\def \C {\textsf{C}}
\def \tp {||}
\def \p {x}
\def \x {z}
\def \V {\textsf{V}}
\def \ls {{\rm LS}}
\def \ma {\Upsilon}
\def \SL {{\rm SL}}
\def \GL {{\rm GL}}
\def \w {\omega}
\def \e {\epsilon}

\numberwithin{equation}{section}

\section{Introduction}\label{sec:intro}
The generalization of the biadjoint scalar amplitude by Cachazo, Early, Guevara and Mizera (CEGM) in \cite{Cachazo:2019ngv} opened the door for a natural generalization of quantum field theory, and a 
lot of exciting progress has occurred in the last few years.
These developments include the identification of the relevant objects
analogous to Feynman diagrams \cite{Borges:2019csl,Cachazo:2019xjx,Cachazo:2022pnx}, 
the formulation of a CEGM stringy version \cite{Arkani-Hamed:2019mrd}, 
and the study of triangulations linked to cluster algebras~\cite{Drummond:2019qjk}.

CEGM amplitudes are defined by generalizing the CHY formula \cite{Cachazo:2013gna,Cachazo:2013hca}, which computes scattering amplitudes for different theories of massless particles in any spacetime dimension. The CHY formula is  an integral over the moduli space ${\cal M}_{0,n}$ of $n$ points on the line $\mathbb{CP}^1$. The moduli space ${\cal M}_{0,n}$ is  the quotient 
$G(2,n)/(\mathbb{C}^*)^n$ of a Grassmannian by its torus action. The CEGM generalization is an integral over the analogous moduli space $X(k,n):=G(k,n)/(\mathbb{C}^*)^n$ for $n$ points in the projective space $\mathbb{CP}^{k-1}$. Note that $X(k,n)$ has
 dimension $(k-1)(n-k-1)$. The poles of CEGM amplitudes correspond to rays of the tropical Grassmannian 
 \cite{speyer2004tropical, speyer2005tropical}.
 
In this paper we focus on the amplitude   $m_n^{(k)}(\mathbb{I},\mathbb{I})$ for the canonical ordering $\mathbb{I}=(123\cdots n)$. We simply write $m_n^{(k)}$
 for this CEGM amplitude. The ordering $\mathbb{I}$ 
   gives rise to a notion of planarity,
   connected to the positive tropical Grassmannian $\textrm{Trop}^+G(k,n)$. 
 In the familiar case  $k=2$, 
 this is the space of planar trees on $n$ leaves, and one recovers the standard partial biadjoint scalar amplitude $m_n^{(2)}$. 
 The CEGM amplitude $m_6^{(3)}$ is displayed explicitly in~(\ref{m36}).

The amplitude $m_n^{(k)}$ can be computed by summing over generalized Feynman diagrams \cite{Borges:2019csl,Cachazo:2019xjx,Cachazo:2022pnx}, which are arrays of trees on $n-k+2$ leaves
\cite[Section 5.4]{MS}. These generalized Feynman diagrams are in correspondence with the maximal cones of the positive tropical Grassmannian $\textrm{Trop}^+G(k,n)$. Therefore, the amplitude $m_n^{(k)}$ can also be computed by using the global Schwinger formula \cite{Cachazo:2020wgu} as a Laplace transform over the whole space $\textrm{Trop}^+G(k,n)$.

The crucial question is: {\em what are CEGM amplitudes actually computing?}
What is their physical meaning?
Since $X(k,n)\simeq X(n-k,n)$, the CEGM amplitude for $k=n-2$ is dual to
the standard $n$-point QFT amplitude $m_n^{(2)}$. The latter is well-studied, and is
   deeply connected to the particle scattering of other theories \cite{Arkani-Hamed:2023swr},
   like the non-linear sigma model (NLSM) or pure Yang-Mills. However, for general $k$ and $n$,
the  CEGM amplitudes are different in nature. 

Knowing their behavior at singularities is important in order to address the question above. For example,  scattering amplitudes in QFT factorize into two lower-point amplitudes at a physical pole. This behavior 
reflects the unitarity and locality constraints on the S-matrix. CEGM amplitudes exhibit exotic factorization behavior at their poles, which highlights the
complicated boundary structure of $X(k,n)$. Nevertheless, it is known that QFT amplitudes emerge from $m_n^{(k)}$ as soft factors in soft limits~\cite{Sepulveda:2019vrz} or as multi-dimensional residues \cite{Cachazo:2022vuo}.

This paper is motivated by recent progress  \cite{Cachazo:2021wsz,Cao:2024gln,Arkani-Hamed:2024fyd,Cao:2024qpp,Zhang:2024iun,Zhou:2024ddy,Zhang:2024efe,Arkani-Hamed:2024nzc},  which has shown that scattering amplitudes in 
QFT and string theory factorize away from poles into lower-point objects after evaluation on certain kinematic subspaces, called {\em split kinematics}. These factorizations are known as splits, and they are related to hidden zeros \cite{Arkani-Hamed:2023swr} which encode fundamental features of scattering amplitudes. These zeros led to new links
   between different theories \cite{Arkani-Hamed:2023swr,Arkani-Hamed:2023jry,Arkani-Hamed:2024nhp} and to novel insights on color-kinematics duality \cite{Bartsch:2024amu} and bootstrap techniques \cite{Rodina:2024yfc}.
 Therefore, to understand the physical implications of the CEGM generalization of QFT,
  it is important to study how $m_n^{(k)}$ decomposes
    not only at  poles but also under  split kinematics.
    
    We here present a splitting of the
CEGM amplitude  $m_n^{(k)}$ into $n-k-1$ factors of the~form
\begin{equation}\label{splitCEGMintro}
 \frac{u_0 + u_1 + \ldots + u_{k-1}}{ u_0 \cdot u_1 \cdot \, \,\ldots\,\,  \cdot u_{k-1}}.
 \end{equation}
This factorization holds on
a split kinematics subspace ${\cal K}^{\textrm{split}}$, which is defined in Section \ref{sec:splitKin}.
Each $u_i$ is an explicit linear combination of generalized kinematic invariants
$\mathsf{s}_I$ where $I \in \binom{[n]}{k}$.

A dual version of $\mathcal{K}^{\rm split}$ appeared
 in recent work of Cachazo and Early. It was called 
{\em conical kinematics} in \cite{Cachazo:2020wgu}
and {\em root kinematics} in \cite{Early:2021solo}. Our paper
also serves as a guide to these articles. In particular, our
results confirm \cite[Conjecture 10.25]{Cachazo:2020wgu}
and \cite[Claim~8.4]{Early:2021solo}.

The rational function (\ref{splitCEGMintro}) arises in the Laurent series for 
the $k$-dimensional beta function
$$
\frac{\Gamma( u_0) \cdot \Gamma(u_1) \cdot \ldots \cdot \Gamma( u_{k-1})}
{\Gamma(\,u_0 + u_1 + \ldots + u_{k-1} \,)} \quad = \quad
(\ref{splitCEGMintro}) \, + \, \hbox{higher order terms}.
$$
Hence (\ref{splitCEGMintro})  is the most basic instance of
a stringy amplitude in the field-theory limit $\alpha' \rightarrow 0$; see
\cite{Arkani-Hamed:2019mrd} and \cite[Example 14]{Sturmfels:2020mpv}.
In Section \ref{sec:simplex} we derive (\ref{splitCEGMintro}) from the
biadjoint scalar amplitude. 

Section \ref{sec:GSF} is devoted to a global Schwinger formula 
for split kinematics. Section  \ref{sec:bicolored}
offers a fresh perspective via tropical determinantal varieties 
and bicolored trees \cite{Dev, MY}. These objects relate the
splitting of CEGM amplitudes  to the matroid amplitudes
introduced by Lam in~\cite{Lam:2024jly}.
The rational function identity
(\ref{eq:tobl1}) =  (\ref{eq:tobl2})
offers a glimpse into a rich theory.

\section{Simplex Amplitudes}\label{sec:simplex}

We begin with the familiar QFT setting. The biadjoint scalar amplitude $m_n^{(2)}$ is 
a rational function in the $\binom{n}{2}$ Mandelstam invariants
$s_{ij}:=(p_i+p_j)^2$, where the $p_i$ are  the momenta of $n$ massless particles.
 The $n$ momentum conservation relations
$\sum_{j=1}^n s_{ij} = 0$ always hold.

Let $\mathcal{K}_\Delta$ denote the $(2n-2)$-dimensional
kinematic subspace given by the invariants
\begin{equation}
\label{eq:simplexkinematics}
 s_{12},s_{13},s_{14},\ldots,s_{1n}\,\,\,{\rm and} \,\, \,s_{23},s_{34},s_{45}, \ldots, s_{n-1,n}
\quad \hbox{together with} \,\,s_{2n}. 
\end{equation}
All other $s_{ij}$ are set to zero. This is an instance of 
{\em minimal kinematics} \cite{MKandPK}. In the setting of
\cite{Early:2024nvf}, it corresponds to the planar $2$-tree
obtained by triangulating the $n$-gon from one vertex.

 The following stringy integral  \cite{Arkani-Hamed:2019mrd} 
 embodies the canonical form of the $(k -1)$-simplex:
\begin{equation}\label{stringy}
    {\cal I}_{\Delta_{k-1}}\,\,=\,\,\,(\alpha')^{k-1}\int_{\mathbb{R}_{\geq0}^{k-1}}\prod_{i=1}^{k-1}\frac{dx_i}{x_i} \,x_i^{\alpha's_{i+2,i+3}} \, \bigl(1\!+\! x_1 \!+\! x_2 \!+ \cdots +\! x_{k-1}   \bigr)^{\alpha's_{2,k+2}} \,.
\end{equation}
Momentum conservation under \textit{simplex kinematics} ${\cal K}_{\Delta}$ yields the following identities:
$$s_{2,k+2}\,+\,\sum_{i=3}^{k+1}s_{i,i+1}\,\,=\,\,s_{2,k+2}+s_{12}\,=\,-s_{23}\,.$$
These imply $s_{2,k+2}\,\,=\,\,-\sum_{i=2}^{k+1}s_{i,i+1}$. This means that on simplex kinematics we can write
\begin{equation}\label{stringysimplex}
    {\cal I}_{\Delta_{k-1}}\,\,=\,\,\,(\alpha')^{k-1}\int_{\mathbb{R}_{\geq0}^{k-1}}\prod_{i=1}^{k-1}\frac{dx_i}{x_i} \,x_i^{\alpha's_{i+2,i+3}} \, \bigl(1\!+\! x_1 \!+\! x_2 \!+ \cdots +\! x_{k-1}   \bigr)^{-\alpha'\sum_{i=2}^{k+1}s_{i,i+1}} \,.
\end{equation}
We define the {\em simplex amplitude}  $m_{\Delta_{k-1}}$ to be the $\alpha'\to0$ limit of (\ref{stringysimplex}). This rational function coincides with the restriction of the standard amplitude
 $m^{(2)}_n$ to  $\mathcal{K}_\Delta$, where now $n=k+2$.

\begin{example}[$k=2$] The $1$-dimensional simplex $\Delta_1$ is a line segment.
Its stringy integral
\begin{equation} \label{m4simplex1}
\!\! \!\!   {\cal I}_{\Delta_1}\,=\,\alpha'  \int_{0}^{\infty}\frac{dx}{x}x^{\alpha's_{34}}\left(1\!+\!x\right)^{-\alpha'(s_{23}+s_{34})} 
 \,=\,  \alpha'\,\frac{\Gamma(\alpha's_{23})\,\Gamma(\alpha's_{34})}{\Gamma(\alpha'(s_{23}+s_{34}))}
\end{equation}
is the Euler beta function, which gives rise to the scattering amplitude of four bosonic open strings \cite{Veneziano:1968yb}.
Here $s_{34}$ and $s_{23}$ are the $s$ and $t$ channels, respectively.  The integral 
in  (\ref{m4simplex1})
converges for $s_{23},s_{34}>0$.
 In the $\alpha'\to0$ limit we obtain the following simple rational function:
\begin{equation} \label{m4simplex}
    {\cal I}_{\Delta_1} 
\, \, \,\, \xrightarrow[\alpha' \to 0]{} \,\,\, \,m_{\Delta_1}
   \, =\,\,\frac{s_{23}+s_{34}}{s_{23}\,s_{34}}.
\end{equation}
The segment amplitude $m_{\Delta_1}$ coincides with the
   $4$-point biadjoint scalar amplitude $m_4^{(2)}$, and with
   (\ref{splitCEGMintro}) for   $u_0 = s_{23}$,  $u_1 = s_{34}$. Note that here no
   kinematic invariant is set to zero on ${\cal K}_{\Delta}$.
      \end{example}

\begin{example}[$k=3$] \label{eq:trianglekinematics}
Triangle kinematics ${\cal K}_{\Delta}$ is given by  $s_{24}=s_{35}=0$. We define
\begin{equation}
\begin{split}
    {\cal I}_{\Delta_2}\,\,= &\,\,\,\,(\alpha')^2\int_{\mathbb{R}_{\geq0}^2}\frac{dx_1}{x_1}\frac{dx_2}{x_2}x_1^{\alpha's_{34}}x_2^{\alpha's_{45}}\left(1+x_1+x_2\right)^{-\alpha'(s_{23}+s_{34}+s_{45})}\,.
    \end{split}
\end{equation}
This  stringy integral, for the canonical form of a triangle, yields the trivariate beta function
$$
    {\cal I}_{\Delta_2}\,\,=\,\,(\alpha')^2\,\frac{\Gamma(\alpha's_{23})\,
    \Gamma(\alpha's_{34})\,\Gamma(\alpha's_{45})}{\Gamma(\alpha'(s_{23}+s_{34}+s_{45}))}.
$$
 Note that ${\cal I}_{\Delta_{2}}$ converges when $s_{23}$, $s_{34}$ and $s_{45}$
are positive. In the $\alpha'\to0$ limit we find
\begin{equation}\label{m5simplex}
\begin{split}
    {\cal I}_{\Delta_2}\,\,\xrightarrow[\alpha' \to 0]{} \,\,\,\,m_{\Delta_2}\,\,=\,\,  \frac{1}{s_{34}s_{45}}+\frac{1}{s_{23}s_{45}}+\frac{1}{s_{23}s_{34}}
    \,\,= \,\, \frac{s_{23}+s_{34}+s_{45}}{s_{23}s_{34}s_{45}}.
    \end{split}
\end{equation}
The triangle amplitude $m_{\Delta_2}$
 is the restriction to  ${\cal K}_{\Delta}$ of the
$5$-point biadjoint scalar amplitude 
\begin{equation}
\label{eq:familiar5}
 m_5^{(2)}\,\,=\,\,\frac{1}{s_{12}s_{34}}+\frac{1}{s_{12}s_{45}}+\frac{1}{s_{23}s_{45}}
 +\frac{1}{s_{15} s_{23}}
 +\frac{1}{s_{15}s_{34}}.
 \end{equation}
For $k=4$, the analogous specialization from $m_6^{(2)}$ to $m_{\Delta_3}$
 appears in \cite[Example~1.5]{Early:2024nvf}. 
{\em Horn uniformization} explains
products of positive linear forms like (\ref{m5simplex})
 and \cite[Theorem~4.1]{Early:2024nvf}.
\end{example}

In general, for $n=k+2$ particles, the stringy integral in (\ref{stringysimplex}) for the canonical form of the $(k-1)$-simplex evaluates to the $k$-dimensional beta function
\begin{equation}
    {\cal I}_{\Delta_{k-1}}=(\alpha')^{k-1}\frac{\Gamma(\alpha's_{23})\Gamma(\alpha's_{34})\cdots\Gamma(\alpha's_{k+1,k+2})}{\Gamma\left(\alpha'\left(s_{23}+s_{34}+\cdots +s_{k+1,k+2}\right)\right)}
\end{equation}
for positive values of the planar variables $s_{i,i+1}$. In the limit $\,\alpha' \rightarrow 0\,$ this yields the following formula, which generalizes
(\ref{m4simplex}) and (\ref{m5simplex}) from $k=2,3$ to arbitrary~$k$:
\begin{equation}
\label{eq:resulting}
        m_{\Delta_{k-1}} \,\, = \,\,\left.m^{(2)}_{k+2}\right|_{{\cal K}_{\Delta}}\,\,=\,\,\,\frac{\sum_{i=2}^{k+1}s_{i,i+1}}{\prod_{i=2}^{k+1}s_{i,i+1}}.
\end{equation}
This is the simplex amplitude
(\ref{splitCEGMintro}), where $u_0,u_1,\ldots,u_{k-1}$ are
adjacent Mandelstam variables.

In conclusion, we have shown that
the $n$-point biadjoint scalar amplitude with kinematics 
$\mathcal{K}_\Delta$ is the leading order in the $\alpha'\to0$ expansion of the stringy  amplitude \eqref{stringysimplex}.
The resulting amplitude (\ref{eq:resulting}) is a rational function
associated to the simplex of dimension $k-1=n-3$.

\section{Positive Parametrization}\label{sec:config}

The CHY formula for QFT scattering amplitudes rests on the
moduli space $\mathcal{M}_{0,n}=X(2,n)$. Likewise, the CEGM amplitude is 
based on the moduli space $X(k,n)$. This space parametrizes configurations of $n$ points in $\mathbb{CP}^{k-1}$.
We represent such configurations by $k \times n$ matrices
\begin{equation} \label{matrix} {\cal X} \quad = \quad \begin{bmatrix}
 0 &  \cdots &\,\, 0 &  \phantom{-}0 & \,\,1 \,\,& \,\,m_{1,1} & m_{1,2} & \cdots & m_{1,n-k} \\
0 &  \cdots &\,\, 0 & -1 &\,\, 0 \,\,&                   \,\,m_{2,1} & m_{2,2} & \cdots & m_{2,n-k} \\
0 & \cdots & \,\,1 & \phantom{-}0 &\,\,0\,\, &                  \,\,   m_{3,1} & m_{3.2} & \cdots & m_{3,n-k} \\
\vdots&\!\reflectbox{$\ddots$}  &\,\, \vdots & \phantom{0} \vdots & \,\,\vdots \,\,&\,\, \vdots & \vdots & \ddots & \vdots \\
(-1)^{k-1} &  \cdots & \,\,0 & \phantom{-} 0 & \,\,0\,\, & \,\,m_{k,1}   & m_{k,2} & \cdots & m_{k,n-k} \\
\end{bmatrix}.
\end{equation}
The $k \times k$ matrix on the left is a signed permutation matrix whose determinant equals $+1$.
Each entry of the right $k \times (n-k)$ matrix $(m_{i,j})$  is a polynomial
in $(k-1) (n-k-1)$ unknowns $x_{r,s}$. These entries are defined recursively.
First, each entry in the first row and column is $1$:
\begin{equation}
\label{Mbasecase}  m_{1,j} = 1 \,\,\,{\rm for}\, \,\, j = 1,2,\ldots,n-k \quad {\rm and} \quad
  m_{i,1} = 1 \,\, \,{\rm for}\, \,\, i = 1,2,\ldots,k. \end{equation}
  Next, each other entry is obtained from the entries above it and to the left of it as follows: 
\begin{equation}
\label{Minduction}
\qquad m_{i,j} \,\, = \,\, m_{i-1,j} \, + \, x_{i-1,1} \cdot \textsf{shift}(m_{i,j-1}) \qquad {\rm for} \,\,\, i,j \geq 2. 
\end{equation}
Here, $\textsf{shift}$ is the operator which replaces the unknown
$x_{i,j}$ by the unknown $x_{i,j+1}$ for all $i,j$.
These rules imply that $m_{i,j}$ is a polynomial
of degree ${\rm min}(i,j)-2$ which is a sum of
$\binom{i+j-2}{i-1}$ monomials. Indeed, if we set $x_{r,s} = 1$
then our $k \times (n-k)$ matrix is just Pascal's triangle.

\begin{example}[$k=2$]
Our matrix representation for the moduli space ${\cal M}_{0,n}=X(2,n)$ is
\begin{equation}
\label{eq:k2} {\cal X} \,\,\, = \,\,\, \begin{bmatrix}
\,\,0 & \,1 & \,\,1\,\, & 1 & \,\,1  &   \,\,1\,\, & \cdots\,\,\,\, \\
\,\,-1 & \,\,\, 0 & \,\,1\,\, & \,1 \!+\! x_{1,1} \,&\, \,\,1+x_{1,1}(1+x_{1,2})\, &\,\,\,  1 + x_{1,1}(1+x_{1,2}(1+x_{1,3})) \,\,\,& \cdots\,\,\,\, \\
\end{bmatrix}.
\end{equation}
For $1 \leq a < b \leq n$, the $2 \times 2$ minor $\mathcal{X}_{ab}$ 
of ${\cal X}$ is an affine-linear form with all coefficients $+1$.
\end{example}

\begin{example}[$k=n-2$]
The moduli space $X(n-2,n)$ is isomorphic to $\mathcal{M}_{0,n}$. We have
\begin{equation} \label{eq:n-2} {\cal X} \quad = \quad \begin{bmatrix}
 0 &  \cdots &\,\, 0 &  \phantom{-}0 & \,\,\,1\, \,\,& \,\,1 \,\, &  \,\, 1 \\
 0 &  \cdots &\,\, 0 & -1 &\,\,\, 0\, \,\,&  \,\,1 \,\, &  1+ x_{1,1}  \\
 0 & \cdots & \,\,1 & \phantom{-}0 &\,\,\,0\,\,\, &    \,\,1 \,\, &   \,\,1+ \sum_{i=1}^2 x_{i,1} \\
  \vdots&\!\reflectbox{$\ddots$}  &\,\, \vdots & \phantom{0} & \vdots & \vdots & \vdots \\
 (-1)^{k-1} &  \cdots & \,\,0 & \phantom{-} 0 &\, \,\,0\,\,\, &  \,\,1 \,\, & \,\,1 + \sum_{i=1}^{n-3} x_{i,1} 
 \end{bmatrix}.
\end{equation}
Each maximal minor of ${\cal X}$ is an affine-linear form in
$n-3$ unknowns with all coefficients $+1$.
\end{example}

\begin{example}[$k=3,n=6$] \label{ex:36}
The space $X(3,6)$ comprises configurations of six points in~$\mathbb{CP}^2$:
\begin{equation} \label{eq:36} {\cal X} \quad = \quad \begin{bmatrix}
\,\, 0 & \, \phantom{-}0 & \,\,\,1\, \,\,& \,\,1 \,\, &  \,\, 1 \,\, & \,\, 1 \,\, \\
\, \, 0 & \,-1 &\,\,\, 0\, \,\,&  \,\,1 \,\, &  1+ x_{1,1}  & 1 \!+\! x_{1,1}(1 \!+\! x_{1,2}) \\
\, \,1 &\, \phantom{-}0 &\,\,\,0\,\,\, &    \,\,1 \,\, &   \,\,1\!+\! x_{1,1}\!+\! x_{2,1}  & m_{3,3}
\end{bmatrix},
\end{equation}
where $\,m_{3,3} \,=\, m_{2,3} + x_{2,1} \cdot {\rm shift}(m_{3,2}) \, = \,
1 \!+\! x_{1,1}(1 \!+\! x_{1,2}) \,+\, x_{2,1} \cdot (1\!+\! x_{1,2}\!+\! x_{2,2}  )$.
Each of the $20$ maximal minors $\mathcal{X}_{ijk}$ of ${\cal X}$ is a sum of monomials in the four unknowns
  $\,x_{1,1},x_{1,2},x_{2,1},x_{2,2}$.
\end{example}

\begin{example}[$k=4,n=9$]
The lower right entry $m_{4,5}$ of the $4 \times 9$-matrix ${\cal X}$ is a quartic.
It is the sum of  $35 = \binom{7}{3}$ monomials of degree $4$
in the $12$ unknowns $x_{ij}$. Explicitly, we have
$$ \begin{matrix} m_{4,5} \,\,\, = \,\,\,
x_{11} x_{12} x_{13} x_{14}+x_{12} x_{13} x_{14} x_{21}
+x_{12} x_{13} x_{14} x_{31}+x_{13} x_{14} x_{21} x_{22}+x_{13} x_{14} x_{22} x_{31}\quad \\
+x_{13} x_{14} x_{31} x_{32}+x_{14} x_{21} x_{22} x_{23}+x_{14} x_{22} x_{23} x_{31}
+x_{14} x_{23} x_{31} x_{32}+x_{14} x_{31}x_{32} x_{33}
+x_{21} x_{22} x_{23} x_{24} \\
+x_{22} x_{23} x_{24} x_{31}+x_{23} x_{24} x_{31} x_{32}
+x_{24} x_{31}  x_{32}  x_{33} + x_{31} x_{32}  x_{33} x_{34}
+x_{11} x_{12} x_{13}+x_{12}x_{13} x_{21} \\
+x_{12} x_{13} x_{31}+x_{13} x_{21} x_{22}+x_{13} x_{22} x_{31}+x_{13} x_{31} x_{32}
+x_{21} x_{22} x_{23}+x_{22} x_{23} x_{31}+x_{23} x_{31} x_{32} \\
+x_{31} x_{32} x_{33} + x_{11} x_{12} +x_{12} x_{21}+x_{12} x_{31}+x_{21} x_{22}
+x_{22} x_{31}+x_{31}x_{32}+
x_{11}+x_{21}+x_{31}+1.
\end{matrix}
$$
Each of the $126$ maximal minors of the matrix ${\cal X}$ is a sum of monomials in the $12$ unknowns.
\end{example}

The matrix ${\cal X}$ gives a birational parametrization $\mathbb{C}^{(k-1) (n-k-1)} \dashrightarrow X(k,n)$
of the moduli space. To see this, note that each parameter $x_{i,j}$ can be expressed as a product of cross~ratios.
For instance, the four parameters for $X(3,6)$ in Example \ref{ex:36} can be recovered as follows:
\begin{equation}
\label{eq:recovery}
 x_{1,1} = \frac{\mathcal{X}_{123} \mathcal{X}_{145}}{\mathcal{X}_{125}\mathcal{X}_{134}},\,\,
     x_{1,2} = \frac{\mathcal{X}_{124} \mathcal{X}_{156}}{\mathcal{X}_{126}\mathcal{X}_{145}},\,\,
     x_{2,1} = \frac{\mathcal{X}_{123}\mathcal{X}_{124} \mathcal{X}_{345}}{\mathcal{X}_{125} \mathcal{X}_{134} \mathcal{X}_{234} },\,\,
x_{2,2} =  \frac{\mathcal{X}_{125} \mathcal{X}_{134} \mathcal{X}_{456} }{\mathcal{X}_{126} \mathcal{X}_{145} \mathcal{X}_{345} }.
\end{equation}

\begin{figure}[h]
\centering
 \includegraphics[width = 6.5cm]{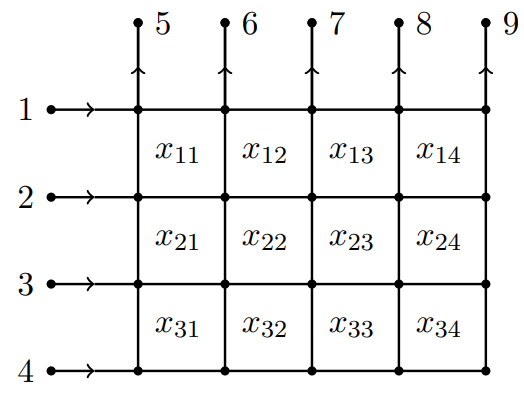}
\vspace{-0.02in}
\caption{Network representation of the matrix $(m_{i,j})$ for $k=4$ and $n=9$.
The polynomial $m_{i,j}$ is a sum over all
north-east paths from node $i$ to node $4+j$. There are $35$ such paths
for $i=4,j=5$.
}
\label{fig:endler}
\end{figure}

Our birational parametrization of $X(k,n)$ is  a variant of the 
{\em network parametrization} which is used widely in
algebraic combinatorics \cite{Drummond:2019qjk, speyer2005tropical}. In physics, this arises in
 the study of the amplituhedron.
The network is a $k \times (n-k)$ grid,  drawn as a directed graph
with source nodes $1,2,\ldots,k$ and sink nodes
$k+1,k+2,\ldots,n$. The regions are labeled
by the parameters $x_{ij}$.
This graph is shown for $k=4$ and $n=9$ in Figure \ref{fig:endler}.
The polynomial $m_{i,j}$ encodes all north-east paths
from node $i$ to node $k+j$. 
There are $\binom{i+j-2}{i-1}$  such paths.
For each path one takes the
product over the labels directly above the horizontal edges.
For instance, the path from $4$ to $9$ which visits the lower right
corner contributes the monomial $x_{31} x_{32} x_{33} x_{34}$.

Our parametrization of the moduli space $X(k,n)$ is
positive in the sense that it
maps the orthant $\mathbb{R}_{>0}^{(k-1)(n-k-1)}$ onto the positive
moduli space \cite{Arkani-Hamed:2020cig}.
Indeed, the Lindstr\"om--Gessel--Viennot Lemma ensures that
each $k \times k$ minor of ${\cal X}$ is a positive integer linear combination of
monomials. So, these minors are positive whenever all $x_{i,j}$ are positive.
Moreover, the $x_{i,j}$ can be expressed as Laurent monomials
in these minors. See (\ref{eq:recovery}) for an illustration.

\section{CEGM Revisited}\label{sec:CEGM}

The $(k,n)$ CEGM amplitude can be defined by the integral representation
\begin{equation} \label{CEGMamplitude}
    m_n^{(k)}\,\,\,=\,\,\int\,\prod_{i=1}^{k-1} \prod_{j=1}^{n-k-1}
    dx_{i,j}\,\,\delta\!\left(\frac{\partial{\cal S}_n^{(k)}}{\partial x_{i,j}}\right)\left(\textrm{PT}_n^{(k)}\right)^2\,.
\end{equation}
We briefly discuss the ingredients in this formula. First, there is the {\em scattering potential}
$${\cal S}_n^{(k)} \,\,\,\,=\sum_{1\leq a_1<a_2<\cdots<a_k\leq n} \!\!\! \mathsf{s}_{a_1,a_2,...,a_k}\,\textrm{log}\,(\mathcal{X}_{a_1,a_2 ,\ldots, a_k}).
$$
Here $\mathcal{X}_{a_1, a_2, \ldots , a_k}$ denotes a maximal minor of the $k \times n$ matrix ${\cal X}$. This is
a positive polynomial in the parameters $x_{ij}$, so the logarithm is well-defined
for positive $x_{ij}$.
 When the parameters $x_{ij}$ are complex,
any branch of the logarithm may be chosen. The coefficients
 $\mathsf{s}_{a_1,a_2,...,a_k}$ are the \textit{generalized kinematic invariants}.
 These are invariant under permuting the indices $a_1,a_2,\ldots,a_k$, and they
 satisfy the masslessness and momentum conservation conditions
 $$\mathsf{s}_{a_1,a_1,a_3,...,a_k}\,=\,0\hspace{13mm}\textrm{and}
 \quad \sum_{1\leq a_2<\cdots<a_k\leq n} \!\!\! \mathsf{s}_{a_1,a_2,...,a_k}
 \,=\,0\,\,\,\,\,{\rm for}\, \,\,\, a_1=1,2,\ldots,n .$$
 The latter relations guarantee that ${\cal S}_n^{(k)}$
  is invariant under scaling the columns of ${\cal X}$.
Hence, the scattering potential is well-defined on the
moduli space $X(k,n)$. The delta functions in \eqref{CEGMamplitude} are distributions that localize the integral to the critical points of the potential ${\cal S}_n^{(k)}$.

The last ingredient for the integral in (\ref{CEGMamplitude}) 
  is the generalized  Parke-Taylor function
$$\textrm{PT}_n^{(k)}\,\,=\,\,\frac{1}{\mathcal{X}_{1,2,\ldots,k}\,\cdot\,\mathcal{X}_{2,3,\ldots,k+1}\,\cdots\,\mathcal{X}_{n,1,\ldots,k-1}}\,.$$
The denominator is the product of the $n$ cyclically adjacent maximal minors of ${\cal X}$.
The CEGM amplitude $m_n^{(k)}$ can be computed by summing over generalized Feynman diagrams \cite{Borges:2019csl,Cachazo:2019xjx,Cachazo:2022pnx}, or by using the global Schwinger formula \cite{Cachazo:2020wgu} as a Laplace transform over $\textrm{Trop}^+G(k,n)$. 
Another method is to sum the values of $({\rm PT}_n^{(k)})^2$ divided by the
toric Hessian of $\mathcal{S}_n^{(k)}$ over all complex solutions of the 
scattering equations $\partial \mathcal{S}_n^{(k)}\!/ \partial x_{i,j} = 0$ for all $i,j$.
The number of solutions grows rapidly.
For instance, it equals $(n-3)!$ for $k=2$ \cite{Cachazo:2013iea}. By \cite{Cachazo:2019ble,Agostini:2021rze}, it is
 $2,26,1272,188112,74570400$ for $k=3$ and $n=5,6,7,8,9$.
 Minimal kinematics \cite{MKandPK, Early:2024nvf} refers to the study of
kinematic subspaces 
where the number of critical points of $\mathcal{S}_n^{(k)}$ drops to one.

The full rational function $m_n^{(k)}$ is very complicated.
It would be desirable to write this amplitude as a sum of reciprocals
of products of linear forms, one for each pole.
Such a representation is available in the $k=2$ case, where it
 mirrors the combinatorics of the associahedron \cite{Arkani-Hamed:2017mur}.
Indeed, the biadjoint scalar amplitude $m_n^{(2)}$
is a sum of terms that correspond to the triangulations of
an $n$-gon. The number of these triangulations is
the Catalan number $\frac{1}{n-1}\binom{2n-4}{n-2}$.
The five terms for $m_5^{(2)}$ 
in (\ref{eq:familiar5}) correspond to the triangulations of a pentagon.

The first exotic instance of CEGM theory is $k=3, n=6$.
For the sake of exposition, we show the explicit formula, which was first computed in \cite{Cachazo:2019ngv}.
   The CEGM amplitude
 $m_6^{(3)}$ is a rational function of degree $-4$ in the $20$ kinematic invariants
 $\mathsf{s}_{ijk}$, which are subject to the momentum conservation constraints
 $ \sum_{j,k} \mathsf{s}_{ijk} = 0$ for $i=1,2,\ldots,6$.
 The amplitude equals 
\begin{equation}
\label{m36}
\begin{matrix}
& \!\! m_6^{(3)}\,=\,\frac{1}{R_{1}\mathsf{s}_{123}\mathsf{s}_{156}\mathsf{s}_{345}}+\frac{1}{\tilde{R}_{2}\mathsf{s}_{123}\mathsf{s}_{156}\mathsf{s}_{345}}+ \frac{1}{\tilde{R}_{1}\mathsf{s}_{126}\mathsf{s}_{234}\mathsf{s}_{456}}+\frac{1}{R_{2}\mathsf{s}_{126}\mathsf{s}_{234}\mathsf{s}_{456}}+\frac{1}{R_1\mathsf{s}_{123}\mathsf{s}_{156}\mathsf{t}_{1234}} \smallskip \\
+ & \frac{1}{\tilde{R}_1\mathsf{s}_{234}\mathsf{s}_{456}\mathsf{t}_{1234}}+\frac{1}{R_2\mathsf{s}_{126}\mathsf{s}_{234}\mathsf{t}_{2345}}+\frac{1}{\tilde{R}_2\mathsf{s}_{156}\mathsf{s}_{345}\mathsf{t}_{2345}}+\frac{1}{R_1\mathsf{s}_{123}\mathsf{s}_{345}\mathsf{t}_{3456}}+\frac{1}{\tilde{R}_1\mathsf{s}_{126}\mathsf{s}_{456}\mathsf{t}_{3456}} \smallskip \\
+ & \frac{1}{R_1\mathsf{s}_{123}\mathsf{t}_{1234}\mathsf{t}_{3456}}+\frac{1}{\tilde{R}_1\mathsf{s}_{456}\mathsf{t}_{1234}\mathsf{t}_{3456}}+\frac{1}{\mathsf{s}_{123}\mathsf{s}_{456}\mathsf{t}_{1234}\mathsf{t}_{3456}}+\frac{1}{\tilde{R}_2\mathsf{s}_{123}\mathsf{s}_{156}\mathsf{t}_{4561}}+\frac{1}{R_2\mathsf{s}_{234}\mathsf{s}_{456}\mathsf{t}_{4561}} \smallskip \\
+ & \frac{1}{\mathsf{s}_{123}\mathsf{s}_{156}\mathsf{t}_{1234}\mathsf{t}_{4561}}+\frac{1}{\mathsf{s}_{156}\mathsf{s}_{234}\mathsf{t}_{1234}\mathsf{t}_{4561}}+\frac{1}{\mathsf{s}_{123}\mathsf{s}_{456}\mathsf{t}_{1234}\mathsf{t}_{4561}}+\frac{1}{\mathsf{s}_{234}\mathsf{s}_{456}\mathsf{t}_{1234}\mathsf{t}_{4561}}+\frac{1}{\tilde{R}_2\mathsf{s}_{156}\mathsf{t}_{2345}\mathsf{t}_{4561}} \smallskip \\
+ & \frac{1}{R_2\mathsf{s}_{234}\mathsf{t}_{2345}\mathsf{t}_{4561}}+\frac{1}{\mathsf{s}_{156}\mathsf{s}_{234}\mathsf{t}_{2345}\mathsf{t}_{4561}}+\frac{1}{\tilde{R}_1\mathsf{s}_{126}\mathsf{s}_{234}\mathsf{t}_{5612}}+\frac{1}{R_1\mathsf{s}_{156}\mathsf{s}_{345}\mathsf{t}_{5612}}+\frac{1}{R_1\mathsf{s}_{156}\mathsf{t}_{1234}\mathsf{t}_{5612}} \smallskip  \\
+ & \frac{1}{\tilde{R}_1\mathsf{s}_{234}\mathsf{t}_{1234}\mathsf{t}_{5612}}+\frac{1}{\mathsf{s}_{156}\mathsf{s}_{234}\mathsf{t}_{1234}\mathsf{t}_{5612}}+\frac{1}{\mathsf{s}_{126}\mathsf{s}_{234}\mathsf{t}_{2345}\mathsf{t}_{5612}}+\frac{1}{\mathsf{s}_{156}\mathsf{s}_{234}\mathsf{t}_{2345}\mathsf{t}_{5612}}+\frac{1}{\mathsf{s}_{126}\mathsf{s}_{345}\mathsf{t}_{2345}\mathsf{t}_{5612}} \smallskip \\
+ & \frac{1}{\mathsf{s}_{156}\mathsf{s}_{345}\mathsf{t}_{2345}\mathsf{t}_{5612}}+\frac{1}{\tilde{R}_1\mathsf{s}_{126}\mathsf{t}_{3456}\mathsf{t}_{5612}}+\frac{1}{R_1\mathsf{s}_{345}\mathsf{t}_{3456}\mathsf{t}_{5612}}+\frac{1}{\mathsf{s}_{126}\mathsf{s}_{345}\mathsf{t}_{3456}\mathsf{t}_{5612}}+\frac{1}{\tilde{R}_2\mathsf{s}_{123}\mathsf{s}_{345}\mathsf{t}_{6123}} \smallskip \\
+ & \frac{1}{R_2\mathsf{s}_{126}\mathsf{s}_{456}\mathsf{t}_{6123}}+\frac{1}{R_2\mathsf{s}_{126}\mathsf{t}_{2345}\mathsf{t}_{6123}}+\frac{1}{\tilde{R}_2\mathsf{s}_{345}\mathsf{t}_{2345}\mathsf{t}_{6123}}+\frac{1}{\mathsf{s}_{126}\mathsf{s}_{345}\mathsf{t}_{2345}\mathsf{t}_{6123}}+\frac{1}{\mathsf{s}_{123}\mathsf{s}_{345}\mathsf{t}_{3456}\mathsf{t}_{6123}} \smallskip \\
+ & \frac{1}{\mathsf{s}_{126}\mathsf{s}_{345}\mathsf{t}_{3456}\mathsf{t}_{6123}}+\frac{1}{\mathsf{s}_{123}\mathsf{s}_{456}\mathsf{t}_{3456}\mathsf{t}_{6123}}+\frac{1}{\mathsf{s}_{126}\mathsf{s}_{456}\mathsf{t}_{3456}\mathsf{t}_{6123}}+\frac{1}{\tilde{R}_2\mathsf{s}_{123}\mathsf{t}_{4561}\mathsf{t}_{6123}}+\frac{1}{R_2\mathsf{s}_{456}\mathsf{t}_{4561}\mathsf{t}_{6123}} \smallskip\\
+ & \frac{1}{\mathsf{s}_{123}\mathsf{s}_{456}\mathsf{t}_{4561}\mathsf{t}_{6123}}+\frac{\mathsf{t}_{1234}+\mathsf{t}_{3456}+\mathsf{t}_{5612}}{R_1 \tilde{R}_1\mathsf{t}_{1234}\mathsf{t}_{3456}\mathsf{t}_{5612}}+\frac{\mathsf{t}_{2345}+\mathsf{t}_{4561}+\mathsf{t}_{6123}}{R_2 \tilde{R}_2\mathsf{t}_{2345}\mathsf{t}_{4561}\mathsf{t}_{6123}}.
\end{matrix}
\end{equation}
Here we use the notation $R_1:=R_{123456}$, $\tilde{R}_1:=R_{125634}$, $R_2:=R_{234561}$ and $\tilde{R}_2:=R_{614523}$ with
$$  R_{abcdef}\,:=\,\mathsf{t}_{abcd}+\mathsf{s}_{cde}+\mathsf{s}_{cdf} \quad {\rm and}
  \quad \mathsf{t}_{abcd}\,:=\,\mathsf{s}_{abc}+\mathsf{s}_{abd}+\mathsf{s}_{acd}+\mathsf{s}_{bcd}. $$
  The amplitude $m_6^{(3)}$ has 16 poles, given by $\mathsf{s}_{123},\mathsf{s}_{234},...,\mathsf{s}_{126}$, $\mathsf{t}_{1234},\mathsf{t}_{2345},...,\mathsf{t}_{6123}$ and $R_1,\tilde{R}_1,R_2$, $\tilde{R}_2$. The number of poles does not match the dimension of the kinematic space, which is $\binom{6}{3}-6=14$. This discrepancy is resolved by the following two identities among the poles:
\begin{equation}
\label{eq:2bipyramids}
\mathsf{t}_{1234}+\mathsf{t}_{3456}+\mathsf{t}_{5612}\,=\,R_{1}+\tilde{R}_{1} \quad {\rm and} \quad
\mathsf{t}_{2345}+\mathsf{t}_{4561}+\mathsf{t}_{6123}\,=\,R_{2}+\tilde{R}_{2}.
\end{equation}
These identities are relevant for the last two of the $48$ summands in \eqref{m36}.
They reflect the two triangulations of a bipyramid, into either two 
or three tetrahedra \cite[Section 2.2.2]{Cachazo:2019ngv}.
According to \cite[Section 5]{speyer2004tropical},
the tropical Grassmannian $\textrm{Trop}\,G(3,6)$ has $15$
 bipyramids  $FFFGG$.
 Precisely two of these $15$ bipyramids lie in the  {\em positive tropical Grassmannian} $\textrm{Trop}^+G(3,6)$.
 The formula (\ref{m36}) reflects the structure of the polytope
 $F_{3,6}$ with f-vector $(16,66,98,48)$, which is derived in \cite[Proposition 6.1]{speyer2005tropical}.
After triangulating each bipyramid into two tetrahedra, given by $R_i + \tilde{R}_i$ for $i=1,2$, one obtains the
{\em $D_4$-associahedron} with f-vector $(16,66,100,50)$.

To appreciate the complexity of the formula (\ref{m36}),
it is instructive to clear denominators. The numerator
is a homogeneous polynomial of degree $12$,
known as  {\em adjoint} in positive geometry.
Its expansion in the $20$ variables
$ \mathsf{s}_{ijk}$ is not unique, because of momentum
conservation. The adjoint is well-defined on the
$14$-dimensional kinematic space. Let us consider the basis
$$ \bigl\{\mathsf{s}_{135}, \mathsf{s}_{136}, \mathsf{s}_{145}, \mathsf{s}_{156}, \mathsf{s}_{235},
 \mathsf{s}_{236},\mathsf{s}_{345}, \mathsf{s}_{146}, \mathsf{s}_{245}, \mathsf{s}_{246},
 \mathsf{s}_{256}, \mathsf{s}_{346}, \mathsf{s}_{356}, \mathsf{s}_{456} \bigr\}. $$
 The unique expansion of our adjoint in these $14$ variables has $1\,552\,098$  monomials.
 This is a hypersurface of degree $12$ in  $\mathbb{P}^{13}$ 
 which deserves to be studied from a geometric perspective.
    
\section{Split Kinematics}\label{sec:splitKin}

Our idea behind the splitting is to replace the
positive moduli space $X^+(k,n)$ by the~polytope
\begin{equation}
\label{Deltaproduct}
 \Delta_{k-1} \times \Delta_{k-1} \times \,\cdots \,\times \Delta_{k-1} . 
 \end{equation}
This has $n-k-1$ factors, so the product of simplices has
dimension $(k-1)(n-k-1)$. The $k$ facet equations of the $j$th factor $\Delta_{k-1}$ are
$x_{1,j}, \,x_{2,j}, \,\ldots,\,x_{k-1,j}\,$ and $\,1 + \sum_{i=1}^{k-1} x_{i,j}$.
We call a Pl\"ucker coordinate $\mathcal{X}_{a_1,a_2,\ldots,a_k}$ 
{\em admissible} if it factors as a product of such facet equations.
The split kinematics space $\mathcal{K}^{\rm split}$ is defined by all
 variables $\mathsf{s}_{a_1,a_2,\ldots,a_k}$ that are admissible.
All kinematic variables indexed by non-admissible tuples $(a_1,a_2,\ldots,a_k)$
 are set to zero on $\mathcal{K}^{\rm split}$.

The number of admissible tuples is found to be $(k+1)(n-k)$.
The $n$ momentum conservation relations remain linearly
independent after setting all $\binom{n}{k} - (k+1)(n-k)$ non-admissible
variables to zero. Hence, the split kinematics space 
$\mathcal{K}^{\rm split}$ has dimension $k(n-k-1)$.

We note that a dual version appeared
in \cite[Section 10.4]{Cachazo:2020wgu}
and in \cite[Section 8]{Early:2021solo}.
The approach by Cachazo and Early differs from ours,
which rests on the facet equations.
The equivalence between the two formulations
yields a proof of \cite[Conjecture 10.25]{Cachazo:2020wgu}.
See~(\ref{eq:splitamplitude}).

\begin{example}[$k=2$] Precisely $3n-6$ of the $\binom{n}{2}$ Mandelstam variables are admissible:
\begin{equation}
\label{eq:admissible2}
s_{12},s_{13},
s_{14},\ldots,s_{1n}, \quad
s_{23},s_{34},s_{45},\ldots,s_{n-1,n}, \quad
s_{24}, s_{35}, s_{46},\ldots, s_{n-2,n}.
\end{equation}
The $n$ Pl\"ucker coordinates
$\mathcal{X}_{12},\mathcal{X}_{13},\ldots,\mathcal{X}_{1n}$ and $\mathcal{X}_{23}$ 
 are equal to~$1$, and
the next $n-3$ Pl\"ucker coordinates are
$\mathcal{X}_{34} = x_{11}$,
$\mathcal{X}_{45} = x_{11} x_{12}$, 
$\mathcal{X}_{56} = x_{11} x_{12} x_{13}$, etc.
The last $n-3$ Pl\"ucker coordinates are
$\mathcal{X}_{24} = 1\!+\!x_{11}$,
$\mathcal{X}_{35} = x_{11}(1\!+\!x_{12})$,
$\mathcal{X}_{46} = x_{11}x_{12}(1\!+\!x_{13})$, etc.
The factors $x_{1j}$ and $1+x_{1j}$
represent the $2n-6$ facets of the cube
$(\Delta_1)^{n-3}$ in (\ref{Deltaproduct}). 

The biadjoint scalar amplitude $m_n^{(2)}$ splits on ${\cal K}^{\textrm{split}}$ into $n-3$ segment amplitudes,
\begin{equation}\label{splitk2}
m_{n,\textrm{split}}^{(2)}\,=\,\left(\frac{1}{s_{12}}+\frac{1}{s_{23}}\right)\left(\frac{1}{s_{123}}+\frac{1}{s_{234}}\right)\left(\frac{1}{s_{1234}}+\frac{1}{s_{2345}}\right)\cdots\left(\frac{1}{s_{n-1,n}}+\frac{1}{s_{n1}}\right)\,,
\end{equation}
where $s_{i_1 i_2 \cdots i_m}:=(p_{i_1}+p_{i_2}+\cdots+p_{i_m})^2$ are the standard planar
kinematic variables. This is the \textit{maximal split} seen in \cite[Section 2.6]{Arkani-Hamed:2024fyd} in the beautiful setting of surfaceology.
The square kinematics $(n=5)$ in
\cite[equation (2.8)]{Arkani-Hamed:2024fyd}
is dual to the triangle kinematics in Example~\ref{eq:trianglekinematics}.

\end{example}

\begin{example}[$k=n-2$]
In this dual scenario, the number of admissible tuples is $2n-2$.
They are $\{1,2,\ldots,n\} \backslash \{n\!+\!1\!-\!i,n\!+\!1\!-\!j\}$
where $s_{ij}$ runs over (\ref{eq:simplexkinematics}). Thus, 
split kinematics for $k=n-2$ is the simplex kinematics for $k=2$ in Section \ref{sec:simplex}.
Indeed, the polytope (\ref{Deltaproduct}) is the simplex $\Delta_{k-1}$, with facets
 $ x_{1,1}, x_{2,1},\ldots,x_{k-1,1}$ and $1 + \sum_{i=1}^{k-1} x_{i,1}$.
Each of these $k$ equations occurs as a unique maximal minor in the
$k \times (k+2)$ matrix ${\cal X}$. In addition, 
$n=k+2$ of the $k \times k$ minors of ${\cal X}$ are equal to $1$.
This accounts for all $k+n$ admissible Pl\"ucker coordinates. 
\end{example}

We next present a detailed study
of split kinematics for the
smallest exotic case.

\begin{example}[$k=3,n=6$]
We think of $X^+(3,6)$ as the product
of two triangles $\Delta_2 \times \Delta_2$.
The kinematic space $\mathcal{K}^{\rm split}$ has dimension $6$.
The $12$ admissible Pl\"ucker coordinates are
$$ \begin{matrix}
 \mathcal{X}_{123} \,=\, \mathcal{X}_{124} \,=\, \mathcal{X}_{125} \,=\, \mathcal{X}_{126} \,=\, \mathcal{X}_{134}
\,=\, \mathcal{X}_{234} \,=\, 1 ,\,\,\,
 \mathcal{X}_{145} =  x_{11}, \,\,\,
\mathcal{X}_{156} = x_{11} x_{12},\, \\
\mathcal{X}_{235} = 1+ x_{11}+x_{21},\,\,\,
\mathcal{X}_{345} =  x_{21},\,\,\,
\mathcal{X}_{346} =  x_{21} (1+x_{12}+x_{22}),\,\,\,
\mathcal{X}_{456} = x_{11} x_{21} x_{22}. 
\end{matrix}
$$
Thus $\mathcal{K}^{\rm split}$ is defined by setting
$\,
\mathsf{s}_{135} \,=\,
\mathsf{s}_{136} \,=\,
\mathsf{s}_{146} \,=\,
\mathsf{s}_{236} \,=\,
\mathsf{s}_{245} \,=\,
\mathsf{s}_{246} \,=\,
\mathsf{s}_{256} =
\mathsf{s}_{356} \,=\, 0$.
Performing this substitution in the CEGM amplitude $m_6^{(3)}$, the expression  \eqref{m36} simplifies to
 \begin{footnotesize}
\begin{equation}\label{eq:36before}
     \left(\frac{\mathsf{s}_{235}}{(\mathsf{s}_{345}{+}\mathsf{s}_{346}{+}\mathsf{s}_{456})(\mathsf{s}_{145}{+}\mathsf{s}_{156}{+}\mathsf{s}_{456})(\mathsf{s}_{145}{+}\mathsf{s}_{156}{+}\mathsf{s}_{235}{+}\mathsf{s}_{345}{+}\mathsf{s}_{346}{+}2\mathsf{s}_{456})}\right) \! \cdot \!
     \left(\frac{\mathsf{s}_{346}}{\mathsf{s}_{156}\mathsf{s}_{456}(\mathsf{s}_{156}{+}\mathsf{s}_{346}{+}\mathsf{s}_{456})}\right).
\end{equation}     
     \end{footnotesize}
This makes some zeroes of the amplitude manifest, namely by further setting $\mathsf{s}_{235}=0$ or $\mathsf{s}_{346}=0$. Momentum conservation on ${\cal K}^{\textrm{split}}$ yields
$\,
\mathsf{t}_{3456}=\mathsf{s}_{345}+\mathsf{s}_{346}+\mathsf{s}_{456}$,
$ \mathsf{t}_{4561}=\mathsf{s}_{145}+\mathsf{s}_{156}+\mathsf{s}_{456}$,
$ \mathsf{t}_{5612}=-(\mathsf{s}_{145}+\mathsf{s}_{156}+\mathsf{s}_{235}+\mathsf{s}_{345}+\mathsf{s}_{346}+2\mathsf{s}_{456})$,
$
\mathsf{s}_{235}=-(\mathsf{t}_{3456}+\mathsf{t}_{4561}+\mathsf{t}_{5612}) $,
and $ \mathsf{s}_{346}=-(\mathsf{s}_{126}+\mathsf{s}_{156}+\mathsf{s}_{456}) $.
Hence, the specialized amplitude (\ref{eq:36before}) also equals
\begin{equation} \label{eq:m36split}
m^{(3)}_{6,\rm split} \,\,= \,\,\,
    \left(\frac{\mathsf{t}_{3456}+\mathsf{t}_{4561}+\mathsf{t}_{5612}}{\mathsf{t}_{3456}\mathsf{t}_{4561}\mathsf{t}_{5612}}\right)\cdot \left(\frac{\mathsf{s}_{126}+\mathsf{s}_{156}+\mathsf{s}_{456}}{\mathsf{s}_{126}\mathsf{s}_{156}\mathsf{s}_{456}}\right).
\end{equation}
Thus, for $k=3,n=6$, the CEGM amplitude is split as the
product of two triangle amplitudes.
\end{example}

We now present our general rule for splitting CEGM amplitudes.
By construction, the restriction of the scattering potential $\mathcal{S}^{(k)}_n$ to the split
kinematic subspace ${\cal K}^{\textrm{split}}$ equals
$$ 
\mathcal{S}^{(k)}_{n,\textrm{split}} \,
= \sum_{j=1}^{n-k-1} \left[
\mathsf{u}_{1,j} \,\textrm{log}(x_{1,j})  + 
\mathsf{u}_{2,j} \,\textrm{log}(x_{2,j})  + \cdots + 
\mathsf{u}_{k-1,j} \,\textrm{log}(x_{k-1,j})  + 
\mathsf{u}_{k,j} \,\textrm{log} \bigl(
1 \!+\! \sum_{i=1}^{k-1} x_{i,j} \bigr)
\right]\!.
$$
Each coefficient $\mathsf{u}_{i,j}$ is the sum of 
a subset of the kinematic variables $\mathsf{s}_{a_1,a_2,...,a_k}$.
We abbreviate the
sum over all $\binom{m}{k}$ kinematic variables coming from an $m$-element subset of the $n$ particles~by
$$ \mathsf{t}_{a_1,a_2,\ldots,a_m} \,\,\,\, = \! \sum_{\{j_1,\ldots,j_k\} \in \binom{[m]}{k}} \!\!\!
\mathsf{s}_{a_{j_1}, \ldots,a_{j_k}}.
$$
The $k \times (n\!-\!k\!-\!1)$ matrix $\bigl( \mathsf{u}_{i,j} \bigr)$ is called the
{\em split kinematics matrix}.  Its entries are

\begin{equation} \label{eq:uformula}
\begin{matrix} &&
\mathsf{u}_{i,j} \, = \,
\mathsf{t}_{j+k-i+1, j+k-i+2,\ldots,k-i-1} & \quad
\hbox{for}\,\,\, i = 1,\ldots,k-1  \\ {\rm and}  &\,\,&
\mathsf{u}_{k,j} \,= \, \mathsf{s}_{j+1,j+2,\ldots,j+k-1,j+k+1}. &
\end{matrix}
\end{equation}
Using momentum conservation under split kinematics $\mathcal{K}^{\rm split}$, we derive the relation
\begin{equation}\label{eq:addu}
    \mathsf{u}_{k,j}\,=\,-\sum_{i=0}^{k-1}\mathsf{u}_{i,j}\,,
\qquad \hbox{with $\,\,\mathsf{u}_{0,j}
\,=\,\mathsf{t}_{j+k+1,j+k+2,...,k-1}$.}
\end{equation}
The quantities $\mathsf{u}_{i,j}$ are poles of the CEGM amplitude $m_n^{(k)}$ unless $i=k$.

The formulas in (\ref{eq:uformula}) are derived by collecting the admissible
summands in the scattering potential $\mathcal{S}^{(k)}_n$.
For our running example $k=3,n=6$, the six entries of the split kinematics matrix are
$\mathsf{u}_{1,1}\,=\,\mathsf{t}_{4561},
\,\mathsf{u}_{2,1}\,=\,\mathsf{t}_{3456},\,\mathsf{u}_{3,1}\,=\,\mathsf{s}_{235}, \,\,\,
\,\mathsf{u}_{1,2}\,=\,\mathsf{s}_{156},\,
\mathsf{u}_{2,2}\,=\,\mathsf{s}_{456},\,\mathsf{u}_{3,2}\,=\,\mathsf{s}_{346}$. 
The two additional poles we defined in (\ref{eq:addu}) are
$\mathsf{u}_{0,1}=\mathsf{t}_{5612}$ and $\mathsf{u}_{0,2}=\mathsf{s}_{126}$.

Split kinematics is always minimal kinematics. This means that the
scattering potential $\mathcal{S}^{(k)}_{n,\textrm{split}} $ has a 
unique critical point $x^*$. The coordinates of this point are the rational functions
$$ x^*_{i,j} \,\, = \,\, - \frac{\mathsf{u}_{i,j}}{ \mathsf{u}_{1,j} + \mathsf{u}_{2,j} + \cdots + \mathsf{u}_{k,j} }. $$
The procedure explained in Section \ref{sec:CEGM} reduces to evaluating
 $({\rm PT}_n^{(k)})^2$ divided by the
toric Hessian of $\mathcal{S}_{n,\textrm{split}}^{(k)}$ at the
unique critical point $x^*$. This yields the split CEGM amplitude
\begin{equation}\label{zerok}
m^{(k)}_{n,\textrm{split}}\,\,=\,\,
\prod_{j=1}^{n-k-1}\frac{\mathsf{u}_{k,j}}{\mathsf{u}_{1,j} \cdot \mathsf{u}_{2,j} \cdot \,
 \,\cdots\, \,\cdot  \mathsf{u}_{k-1,j} \cdot \left(\sum_{i=1}^k\mathsf{u}_{i,j}\right)}. 
 \end{equation}
This is equation \eqref{eq:36before} when $k=3,n=6$. Equation \eqref{zerok} makes manifest that the amplitude vanishes if we further set any $\mathsf{u}_{k,j}$ to $0$.
Momentum conservation \eqref{eq:addu} now implies that the split CEGM amplitude has 
precisely the form promised towards the end of the Introduction:
\begin{equation}
\label{eq:splitamplitude}
m^{(k)}_{n,\rm split} \,\,\,= \,\,\,
\prod_{j=1}^{n-k-1} \frac{ \mathsf{u}_{0,j} + \mathsf{u}_{1,j} + \cdots + \mathsf{u}_{k-1,j}}{\mathsf{u}_{0,j} \cdot \mathsf{u}_{1,j} \cdot \, \,\cdots\, \,\cdot  \mathsf{u}_{k-1,j}}.
\end{equation}
This is the dual formula in  \cite[Conjecture 10.25]{Cachazo:2020wgu}.
It matches the  product structure in (\ref{Deltaproduct}).
The case $k=3,n=6$ was seen in (\ref{eq:m36split}).
The split amplitudes for $k=2$ and $n=k-2$ are
derived from the discussion below  (\ref{eq:admissible2}).
These also show that our splitting
breaks the symmetry between $X(k,n)$ and $X(n-k,n)$.
Split kinematics for $X(k,n)$ rests on the product
of $n-k-1$ simplices of dimension $k-1$, while
split kinematics for $X(n-k,n)$ rests on the product
of $k-1$ simplices of dimension $n-k-1$.
The split kinematics matrix for $X(n-k,n)$ has
format $(n-k) \times (k-1)$. In particular, this matrix is not the
transpose of that for~$X(k,n)$. 

We point out that \eqref{eq:splitamplitude} is not the $k$-analog of a maximal split, in the sense of \cite[Section 2.6]{Arkani-Hamed:2024fyd}, as it is not a product of segment amplitudes. We conclude this section with a question that connects
two  recent threads on scattering amplitudes:
Is there a   {\em surfaceology for CEGM}?
  
\section{Global Schwinger Formula}\label{sec:GSF}

We here discuss a global Schwinger formula
for the split CEGM amplitude. This is a Laplace integral
over a Bergman fan, so it is a special case of
Lam's construction in \cite[Section 10.4]{Lam:2024jly}.

We write ${\cal D}_k^{p\times q}$ for the variety of $p\times q$ matrices of rank $\leq k$,
and we write ${\cal T}_k^{p\times q}$ for the tropicalization of this determinantal variety.
Throughout this section we fix $p=k+1$ and $q=n-k-1$. 
The maximal minors
form a {\em tropical basis} of $\mathcal{D}_k^{p \times q}$, by \cite[Theorem 5.3.25]{MS}.
This means that ${\cal T}_k^{p\times q}$ 
consists of all $p \times q$ matrices such that the
minimum is attained twice for each of its tropical minors of size
$p \times p$. We shall use ${\cal T}_k^{p\times q}$ to construct our amplitudes.


We obtain a positive parametrization by
 multiplying a $p\times k$ matrix with a $k\times q$ matrix:
 \begin{equation}\label{detparam}
\!\!\begin{bmatrix}  
1 & 0 & 0 & \cdots & 0 \\  
0 & 1 & 0 & \cdots & 0 \\ 
\vdots & \vdots & \vdots & \ddots & \vdots \\
0 & 0 & 0 & \cdots & 1 \\
1 & 1 & 1 & \cdots & 1 \\  
\end{bmatrix}  
\begin{bmatrix}  
1 & 1 & \cdots & 1 \\  
x_{1,1} & x_{1,2} & \cdots & x_{1,n-k-1} \\ 
x_{2,1} & x_{2,2} & \cdots & x_{2,n-k-1} \\ 
\vdots & \vdots & \ddots & \vdots \\ 
x_{k-1,1} & x_{k-1,2} & \cdots & x_{k-1,n-k-1} \\
\end{bmatrix}  
\, =\,
\begin{bmatrix}  
1 &  \cdots & 1 \\
x_{1,1} & \cdots & x_{1,n-k-1} \\  
x_{2,1} & \cdots & x_{2,n-k-1} \\  
\vdots  & \ddots & \vdots \\  
x_{k-1,1}  & \cdots & x_{k-1,n-k-1} \\  
1\!+\!\sum_{i=1}^{k-1}x_{i,1} &  \cdots & 1\!+\!\sum_{i=1}^{k-1}x_{i,n-k-1} \\ 
\end{bmatrix}\!  .
\end{equation}
Here $x_{i,j}\in\mathbb{R}^+$. This linear map extends to a parametrization of $\mathcal{D}_k^{p \times q}$
by the torus action that scales the rows and columns.
Therefore, the variety of maximal minors is parametrized by linear forms times monomials, so
the results in \cite[Section 5.5]{MS} apply. They imply that
$\mathcal{T}_k^{p \times q}$ has the structure of a Bergman fan.
See Figure \ref{pic3x3} for the case $k=2$, $p=q=3$.

What follows is 
inspired by the global Schwinger parametrization on the positive tropical Grassmannian \cite{Cachazo:2020wgu} 
and by the construction of amplitudes from matroids in \cite{Lam:2024jly}.
The \textit{tropical scattering potential} on the tropical 
determinantal variety ${\cal T}_k^{p\times q}$ is the function
\begin{equation} \label{eq:tropF0}
    F_{p,q}^{(k)}(x) \quad :=\sum_{\substack{i\in\{1,2,...,p\} \\ j\in\{1,2,...,q\}}}\mathsf{u}_{i-1,j}\,\,\,\mathcal{X}_{i,j}^{\textrm{trop}}\,,
\end{equation}
where $\mathcal{X}_{i,j}^{\textrm{trop}}$ are the tropicalized entries of the matrix \eqref{detparam}. 
The coefficients $\mathsf{u}_{i,j}$ are kinematic variables.
The coefficient $\mathsf{u}_{0,j}$ does not explicitly enter into the tropical potential since $\mathcal{X}^{\textrm{trop}}_{1,j}=0$ for all $j$. 
Note that (\ref{eq:tropF0}) is
the tropical version of the split CEGM scattering potential ${\cal S}^{(k)}_{n,\textrm{split}}$ seen in Section \ref{sec:splitKin}. Explicitly, the tropical potential is given by the piecewise-linear formula
\begin{equation}\label{tropF}
    F_{p,q}^{(k)}(x)\,\,\, =\,\,\,
    \sum_{j=1}^{q}\left[\,\sum_{i=1}^{p-2}\mathsf{u}_{i,j}\, x_{i,j}+\mathsf{u}_{k,j} \cdot \textrm{min}(0,x_{1,j},x_{2,j},...,x_{k-1,j})\right].
\end{equation}
Here, the $x_{i,j}$ are tropical variables, 
and $(u_{i,j})$ is the split kinematic matrix from Section \ref{sec:splitKin}.

 We define the following integral as a Laplace transform over 
 the tropical variety ${\cal T}_k^{p\times q}$:
\begin{equation}\label{eq:laplace}
    I_{p,q}\,\,=\,\,\int_{\mathbb{R}^{(p-2)q}}d^{(p-2)q}x\,\,\,\textrm{exp}(-F_{p,q}^{(k)}(x))\,.
\end{equation}
This integral decomposes into the product of $q=n-k-1$ independent integrals of the form
\begin{equation}\label{eq:gsfsimplex}
    \int_{\mathbb{R}^{p-2}}d^{p-2}x
    \,\textrm{exp}(-F_{k+2,j})\,\,=\,\,\frac{\sum_{i=0}^{k-1} \mathsf{u}_{i,j}}{\prod_{i=0}^{k-1} \mathsf{u}_{i,j}}\,,
\end{equation}
with $F_{k+2,j}:=\sum_{i=1}^{k-1}\mathsf{u}_{i,j}\, x_{i,j}+\mathsf{u}_{k,j}\,\textrm{min}(0,x_{1,j},x_{2,j},...,x_{k-1,j})$. 
To obtain the rational function on the right hand side of \eqref{eq:gsfsimplex}, we have used the following
momentum conservation equations:
\begin{equation}\label{momck}
 \qquad   \sum_{i=0}^{p-1}\mathsf{u}_{i,j}\,=\,0\hspace{10mm}{\rm for} \,\,\, j\in\{1,2,\ldots,q\}.
\end{equation}

We conclude that \eqref{eq:laplace} is a Laplace integral over a Bergman fan, and it evaluates to
\begin{equation}\label{splitCEGM}
    I_{p,q} \,\,=\,\,\prod_{j=1}^{q} \left(\frac{\sum_{i=0}^{p-2} \mathsf{u}_{i,j}}{\prod_{i=0}^{p-2} \mathsf{u}_{i,j}}\right).
\end{equation}
The coefficients $\mathsf{u}_{i,j}$ are the entries of the \textit{enlarged} split kinematics matrix, defined as in \eqref{eq:uformula}, but now with the $\mathsf{u}_{0,j}$.
This implies that \eqref{momck} are the momentum conservation conditions under split kinematics $\mathcal{K}^{\rm split}$. In fact, $F_{k+2,j}$ is the tropical potential for the $(k+2)$-point biadjoint scalar amplitude with kinematics ${\cal K}^{\textrm{split}}$. Therefore, equation \eqref{eq:gsfsimplex} is the simplex amplitude $m_{\Delta_{k-1}}$, and the integral $I_{p,q}$ matches our split CEGM amplitude $m^{(k)}_{n,\rm split}$ in \eqref{eq:splitamplitude}.

We conclude this section with another explicit example. This will illustrate
  how our split kinematics breaks the symmetry between $X(k,n)$ and $X(n-k,n)$ for $k>2$, and how this leads to another split kinematics subspace, namely the {\em root kinematics} defined by Early \cite{Early:2021solo}.

\begin{example} $X(3,7)\simeq X(4,7)$. The kinematics ${\cal K}^{\textrm{split}}$ for $X(3,7)$ sets to zero all $\mathsf{s}_{abc}$ except for the $16$ in $\{\mathsf{s}_{i,i+1,i+2},\mathsf{s}_{124},\mathsf{s}_{125},\mathsf{s}_{126},\mathsf{s}_{134},\mathsf{s}_{145},\mathsf{s}_{156},\mathsf{s}_{235},\mathsf{s}_{346},\mathsf{s}_{457}\}$.
This yields the splitting
   \begin{equation}
       m^{(3)}_{7,\textrm{split}}=\left(\frac{\mathsf{t}_{56712}+\mathsf{t}_{45671}+\mathsf{t}_{34567}}{\mathsf{t}_{56712}\,\mathsf{t}_{45671}\,\mathsf{t}_{34567}}\right)\left(\frac{\mathsf{t}_{6712}+\mathsf{t}_{5671}+\mathsf{t}_{4567}}{\mathsf{t}_{6712}\,\mathsf{t}_{5671}\,\mathsf{t}_{4567}}\right)\left(\frac{\mathsf{s}_{712}+\mathsf{s}_{671}+\mathsf{s}_{567}}{\mathsf{s}_{712}\,\mathsf{s}_{671}\,\mathsf{s}_{567}}\right).
   \end{equation}
   The kinematics ${\cal K}^{\textrm{split}}$ for $X(4,7)$ sets to zero all kinematic variables $\mathsf{s}_{abcd}$ except for the $15$ in $\{\mathsf{s}_{i,i+1,i+2,i+3},\mathsf{s}_{1235},$ $\mathsf{s}_{1236},\mathsf{s}_{1245},\mathsf{s}_{1256},\mathsf{s}_{1345},\mathsf{s}_{1456},\mathsf{s}_{2346},\mathsf{s}_{3457}\}$. 
   This produces the splitting
   \begin{equation}
       m^{(4)}_{7,\textrm{split}}=\left(\frac{\mathsf{t}_{67123}+\mathsf{t}_{56712}+\mathsf{t}_{45671}+\mathsf{t}_{34567}}{\mathsf{t}_{67123}\,\mathsf{t}_{56712}\,\mathsf{t}_{45671}\,\mathsf{t}_{34567}}\right)\left(\frac{\mathsf{s}_{7123}+\mathsf{s}_{6712}+\mathsf{s}_{5671}+\mathsf{s}_{4567}}{\mathsf{s}_{7123}\,\mathsf{s}_{6712}\,\mathsf{s}_{5671}\,\mathsf{s}_{4567}}\right).
   \end{equation}
Geometrically, $m_{7,\textrm{split}}^{(3)}$ degenerates into the product of three triangles while $m_{7,\textrm{split}}^{(4)}$ degenerates into the product of two tetrahedra. By dualizing ${\cal K}^{\textrm{split}}$ in these two cases, we can also split $m_{7,\textrm{split}}^{(3)}$ into the product of two tetrahedra and $m_{7,\textrm{split}}^{(4)}$ into the product of three triangles.

\end{example}

\section{Amplitudes from Determinantal Varieties}\label{sec:bicolored}

In the previous section, split CEGM amplitudes were derived from
a linear parametrization (\ref{detparam}) of the variety of maximal minors.
We now turn to a different determinantal variety,
namely matrices of rank two.
This variety has a linear structure as well. Its tropicalization 
is also a Bergman fan: it is
the space of bicolored trees \cite{Dev, MY}.
We here introduce that space  to the physics community,
and we argue that (\ref{eq:T2pq}) leads to a new class of amplitudes.
The primary aim of this section is to suggest
future research in the context of \cite[Problem 2.6]{Lam:2024jly}.

The Grassmannian $G(2,n)$ is a manifold of
dimension $2n-4$ which parametrizes 
vector subspaces
of dimension $2$ in $\mathbb{C}^n$. 
This manifold is a subvariety 
 of the projective space $\mathbb{P}^{\binom{n}{2}-1}$. In this embedding, each
  point of $G(2,n)$ is a skew-symmetric $n \times n$ matrix
$V = (v_{ij})$ of rank $2$, whose entries $v_{ij}$
are the Pl\"ucker coordinates.
The vector subspace is the image of~$\,V$.

Fix integers $p \geq 2$ and $q \geq 2$ such that $p + q = n$.
Let $\mathcal{D}^{p \times q}_2$ denote the projective
 variety of all $p \times q$ matrices of rank $\leq 2$.
This variety has dimension  $2n-5=2p+2q-5$. 

Every skew-symmetric $n \times n$ matrix $V$ can be written in the block form
 \begin{equation} \label{eq:Umatrix}
V \,\, = \,\, \begin{bmatrix} \,\,P & X\, \,\\ - X^T & Q\, \, \end{bmatrix} ,
\end{equation} 
where $P$ and $Q$ are skew-symmetric matrices of size
$p \times p$ and $q \times q$, respectively, and $X$
is an arbitrary $p \times q$ matrix. If $V$ has rank $2$
then $X$ has rank $\leq 2$. This defines a rational map
\begin{equation}
\label{eq:GrMap} G(2,n) \,\dashrightarrow \,\mathcal{D}^{p \times q}_2\,
 \subset\, \mathbb{P}^{pq-1}\,,\,\, V \,\mapsto \, X .
\end{equation}

\begin{prop}
    The map (\ref{eq:GrMap}) is surjective. 
Over a rank $\,2$ matrix $X \in \mathcal{D}^{p \times q}_2$, the fiber of~(\ref{eq:GrMap}) is the curve
consisting of all matrices
\begin{small} $  \begin{bmatrix} \,\,\,tP & X\, \,\\ - X^T &\! t^{-1}Q\, \,\end{bmatrix} $,\end{small}
where $t \in \mathbb{P}^1$. Here, $P \in G(2,p)$ is recovered as
 the column space of $\,X$, and
$\,Q \in G(2,q)$ is recovered as the row space of~$\,X$.
\end{prop}

\begin{example}[$p=q=2$]
Consider the smallest possible case, when $V$ is the $4 \times 4$ matrix
 \begin{small} $$ V \,\, = \,\,
 \begin{bmatrix}
\,\,0 & p_{12} & x_{11} & x_{12} \\
-p_{12} & 0 & x_{21} & x_{22} \\
-x_{11} & -x_{21} & 0 & q_{12} \\
-x_{12} & -x_{22} & -q_{12} & 0 
\end{bmatrix}.$$  \end{small}
 The $2 \times 2$ matrix $X = (x_{ij})$
 is a given point in $\mathcal{D}^{2 \times 2}_2 = \mathbb{P}^3$. We require
  ${\rm rank}(V) \leq 2$, so  the Pfaffian of $V$ is zero. This is equivalent to 
 $\, p_{12} \cdot q_{12}  \, = \, x_{11} x_{22} - x_{12} x_{21} \,= \, {\rm det}(X) $.
 Hence $p_{12}$ and $q_{12}$
 can be recovered up to scale from the determinant of $X$.
 Geometrically, the general fiber of the map
 $G(2,4) \rightarrow \mathbb{P}^3$ is the closure $\mathbb{P}^1$ of the torus $\mathbb{C}^*$.
 This generalizes to $p,q \geq 2$.
\end{example}

The torus $(\mathbb{C}^*)^n = (\mathbb{C}^*)^p \times (\mathbb{C}^*)^q$
acts on the domain and the range of (\ref{eq:GrMap}),
and this map is equivariant with respect to these torus actions.
The quotient of $G(2,n)$ modulo $(\mathbb{C}^*)^n$ is the
moduli space $\mathcal{M}_{0,n}$.
We obtain the same moduli space as the quotient of the image.
However, the tropical compactifications \cite[Section 6.4]{MS} of these
two copies of  $\mathcal{M}_{0,n}$ are different.
We shall see that (\ref{eq:GrMap}) gives rise to a
piecewise-linear map between two different tree spaces.

The tropical Grassmannian $\mathcal{G}_n := {\rm trop}\,G(2,n)$
   parametrizes metric trees on $n$ leaves \cite[Section 4.3]{MS}.
   Using max-plus convention, the tropical Pl\"ucker coordinate
$d_{ij} = -{\rm val}(v_{ij})$ is the distance between leaf $i$ and leaf $j$
in the tree. Combinatorially, this tree space is a simplicial
complex of dimension $n-4$ with  $2^{n-1} -n-1$ vertices and
$(2n-5)!! = 1 \cdot 3 \cdot 5 \cdot \ldots \cdot (2n-5)$ facets.
For instance, if $n=5$ then this is the Petersen graph,
with $10$ vertices and $15$ edges.

The articles  \cite{Dev, MY} studied 
 the tropicalization of the variety of rank $2$ matrices. This is
 \begin{equation} \label{eq:T2pq}
 \mathcal{T}_2^{p \times q} \,\, := \,\, {\rm trop}(\mathcal{D}^{p \times q}_2). 
 \end{equation}
 Points in $\mathcal{T}_2^{p \times q}$ can be viewed as metric trees with $p$ leaves
that are marked with $q$ points. The role of $p$ and $q$ is symmetric, so the points in
$\mathcal{T}_2^{p \times q}$ are also metric trees with $q$ leaves
that are marked with $p$ points. 
Develin explains this identification in \cite[Section 3]{Dev}.
Markwig and Yu \cite[Section 2.3]{MY} work with  the realization of
$\mathcal{T}_2^{p \times q}$ as a subfan of the usual
tree space $\mathcal{G}_n$. Namely, this is the subfan of
{\em bicolored trees}.
Such trees for $p=q=3$ are shown below (\ref{eq:tobl1}).

The map (\ref{eq:GrMap}) arises from a coordinate projection, and hence so does its tropicalization
\begin{equation}
\label{eq:TrMap1} \mathcal{G}_n \,\,\rightarrow \,\, \mathcal{T}_2^{p \times q}. 
\end{equation}
In tropical Pl\"ucker coordinates, written analogously to (\ref{eq:Umatrix}),
the map (\ref{eq:TrMap1}) is the projection from the space of
symmetric $n \times n$ matrices (without diagonal)
to the space of $p \times q$ matrices. It  takes any
tree metric $(d_{ij})$ to the
$p \times q$ matrix whose entries are the bipartite distances
$d_{ij}$ for $i=1,\ldots,p$ and $j = p+1,\ldots,n$.
Combinatorially, we start from a metric tree with $n = p+q$
leaves, and we retract $q$ of the leaves onto the tree
spanned by the other $p$ leaves. 

\begin{example}[$p=q=3$]\label{ex33}
The tree space $\mathcal{T}_2^{3 \times 3}$ is defined by the
tropical $3 \times 3$ determinant
$$ \begin{matrix}
    \textrm{tropdet}(X ) & =  & 
   \textrm{min}(\,x_{11}+x_{22}+x_{33},\,x_{11}+x_{23}+x_{32},\,x_{12}+x_{21}+x_{33}, \\ & & \qquad\,\,\,
  \,x_{12}+x_{23}+x_{31},\,x_{13}+x_{21}+x_{32} ,  x_{13}+x_{22}+x_{31}\,).
\end{matrix}
$$
The minimum being attained twice specifies a union of $\binom{6}{2}=15$ convex polyhedral cones:
$$ \begin{matrix}
 & x_{11}+x_{22}+x_{33}=x_{11}+x_{23}+x_{32} \,\leq \,\textrm{min}(x_{12}+x_{21}+x_{33},\ldots,x_{21}+x_{32}+x_{13})\\ \textrm{or} \hspace{3mm} & x_{11}+x_{22}+x_{33}=x_{12}+x_{21}+x_{33}\,\leq \,\textrm{min}(x_{11}+x_{23}+x_{32},\ldots,x_{21}+x_{32}+x_{13})\\ 
\textrm{or} \hspace{3mm} & \cdots \quad  \cdots \quad \cdots \quad \cdots  \qquad \qquad
\qquad \qquad \cdots \quad \cdots \quad \cdots \quad \cdots \qquad \qquad \\
 \textrm{or} \hspace{3mm} & x_{12}+x_{23}+x_{31}=x_{21}+x_{32}+x_{13} \,\leq \,\textrm{min}(x_{11}+x_{22}+x_{33},\ldots,x_{13}+x_{22}+x_{31}).
 \end{matrix}
 $$
This is a fan
over a $2$-dimensional polyhedral complex with
$9$ vertices, $18$ edges and $15$ facets,
as shown in Figure \ref{pic3x3}.
 By \cite{MY}, each facet corresponds to a bicolored
tree, with $3+3$ leaves.

\begin{figure}[h!]
	\centering
	\includegraphics[width=0.7\linewidth]{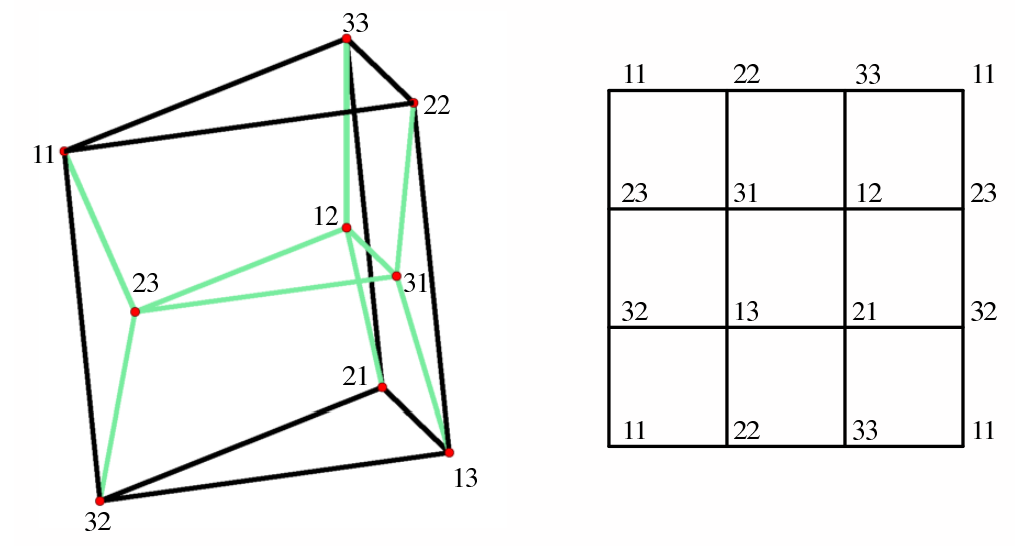}
	\vspace{-0.2in} \caption{
	The tropical variety defined by the $3\times3$ determinant. Image taken from \cite[Figure 3.5]{Berndbook}.}
	\label{pic3x3}
\end{figure} 

These trees can be understood from the map
 (\ref{eq:TrMap1}) for $p=q=3$.
 Recall (e.g.~from \cite[Example 4.3.15]{MS})
that the tree space $\mathcal{G}_6$ is the cone over a $2$-dimensional
simplicial complex with $25$ vertices, $105$ edges and $105$ triangles.
Each triangle corresponds to a trivalent tree with $6$ leaves, which is
encoded by a  triple of splits. These correspond to propagators in physics.
  Among the $105$ trivalent trees
that comprise $\mathcal{G}_6$,
there are $15$ {\em snowflake trees},
encoded by a triple of splits like $\{\{1,2\}, \{3,4\}, \{5,6\} \}$,
and $90$ {\em caterpillar trees}, encoded by a triple like
$\{\{1,2\}, \{1,2,3\}, \{5,6\} \}$.
The $25=15+10$ vertices of $\mathcal{G}_6$ are the splits,
like $\{i,j\}$ or $\{i,j,k\}$.

The map (\ref{eq:TrMap1}) takes six of the snowflake triangles
onto the six triangles in $\mathcal{T}_2^{3 \times 3}$.
These snowflakes are indexed by three splits  $\{i,j\}$
where $i \in\{1,2,3\} $ and $j \in \{4,5,6\}$.
The other nine snowflakes are contracted to the nine
vertices of $\mathcal{T}_2^{3 \times 3}$. 
Among the $90$ caterpillar triangles, $81$ triangles are mapped
into the nine squares of $\mathcal{T}_2^{3\times 3}$.
 Here the preimage
of each square consists of $9$ caterpillar triangles in $\mathcal{G}_6$.
The remaining $9$ caterpillar triangles are contracted to the cone
point of $\mathcal{G}_6$.
These are the caterpillars whose middle split is $\{1,2,3\}$ or $\{4,5,6\}$.
\end{example}

For applications to amplitudes, the positive part of
 $\mathcal{T}_2^{p \times q}$ is especially relevant.
 This tropical space was studied in~\cite{BLS}.  
 The positive part of the map (\ref{eq:TrMap1}) gives a parametrization. But,
 it  depends crucially
on the labeling of the rows and columns of the matrix
$V$. In the example above, the labels
$1,2,3$ referred to the matrix $P$ and the
labels $4,5,6$ referred to $Q$. However, 
to match the gauge fixing for $\mathcal{M}_{0,6}$ commonly
 used in the physics literature,
one takes $1,3,5$ for $P$ and $2,4,6$ for $Q$.
This makes a big difference when it comes to 
positive tree space. In other words, for a tree
in $\mathcal{G}_6$ to be planar depends on which labeling is used.

In Lam's theory \cite{Lam:2024jly},
the tropical variety $\mathcal{T}_2^{p \times q}$ is the
Bergman fan of the graphic matroid of the
complete bipartite graph $K_{p,q}$.
The labelings 
correspond to the choice of a~{\em tope} for this oriented matroid.
And, the tope determines the amplitude, by
\cite[Section 17]{Lam:2024jly}. The global Schwinger
formula for that amplitude is the Laplace transform
of the Bergman fan~$\mathcal{T}_2^{p \times q}$.

What we are proposing  is a new class of amplitudes, where
bicolored trees play the role of Feynman graphs, and tree
space $\mathcal{G}_n = {\rm trop}(\mathcal{M}_{0,n})$ is replaced
by its image under~(\ref{eq:TrMap1}).
These new amplitudes are integrals over a very affine variety,
as suggested in \cite[Problem 2.6]{Lam:2024jly},
 namely the determinantal variety
$\mathcal{D}^{p \times q}_2 $ modulo its torus~$ (\mathbb{C}^*)^n$.
In other words, in the derivation of the
biadjoint scalar amplitude $m^{(2)}_n$ in
\cite[Section 19.1]{Lam:2024jly},
the complete graph $K_n$ is now replaced by the bipartite graph $K_{p,q}$.
We conclude with an illustration for $p=q=3$.

\begin{example}[Triangular prism] 
\label{ex:toblerone}
The theory described above yields the following amplitude:
\begin{equation}
\label{eq:tobl1} \frac{1}{u_{\textcolor{blue}{1}\textcolor{red}{1}} u_{\textcolor{blue}{2}\textcolor{red}{2}} u_{\textcolor{blue}{3}\textcolor{red}{3}}} \,+\,
     \frac{1}{u_{\textcolor{blue}{1}\textcolor{red}{3}} u_{\textcolor{blue}{2}\textcolor{red}{1}} u_{\textcolor{blue}{3}\textcolor{red}{2}}} \,+\,
\frac{u_{\textcolor{blue}{1}\textcolor{red}{1}}+u_{\textcolor{blue}{1}\textcolor{red}{3}}}{u_{\textcolor{blue}{1}\textcolor{red}{1}} u_{\textcolor{blue}{1}\textcolor{red}{3}} u_{\textcolor{blue}{2}\textcolor{red}{2}} u_{\textcolor{blue}{3}\textcolor{red}{2}}} \,+\, 
\frac{u_{\textcolor{blue}{2}\textcolor{red}{1}}+u_{\textcolor{blue}{2}\textcolor{red}{2}}}{u_{\textcolor{blue}{1}\textcolor{red}{3}} u_{\textcolor{blue}{2}\textcolor{red}{1}} u_{\textcolor{blue}{2}\textcolor{red}{2}} u_{\textcolor{blue}{3}\textcolor{red}{3}}} \,+\,
\frac{u_{\textcolor{blue}{3}\textcolor{red}{2}}+u_{\textcolor{blue}{3}\textcolor{red}{3}}}{u_{\textcolor{blue}{1}\textcolor{red}{1}} u_{\textcolor{blue}{2}\textcolor{red}{1}} u_{\textcolor{blue}{3}\textcolor{red}{2}} u_{\textcolor{blue}{3}\textcolor{red}{3}}}.
\vspace{-0.8cm}
\end{equation}
\begin{figure}[h!]
\hspace{11mm}\includegraphics[width=0.8\linewidth]{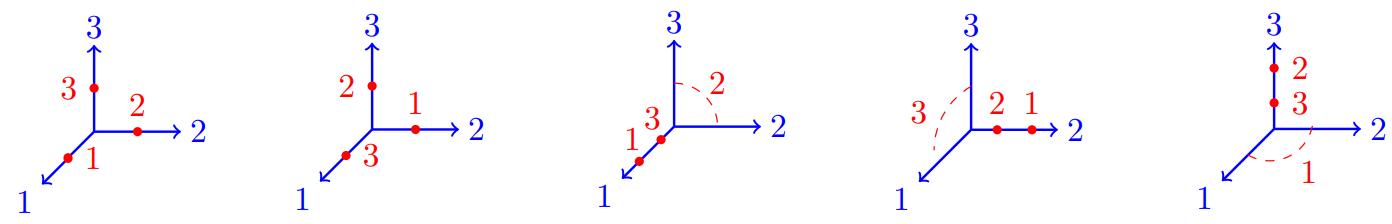}
	\label{bicoloredtrees}
\end{figure} 

Each of the five summands corresponds to the bicolored tree
drawn below it. These trees label the two triangle and three square facets of the
triangular prism with black edges on the left of  Figure \ref{pic3x3}.
We identify $\mathcal{T}^{3\times 3}_2$ with the
Bergman fan of the {\em cographic matroid} of $K_{3,3}$.
This is the tropicalization of the space of
$3 \times 3$ matrices with zero row sums and zero column sums.
The amplitude (\ref{ex:toblerone}) equals that in \cite{Lam:2024jly} if we choose the
tope given by the  sign pattern
$$ \qquad \qquad \qquad  \qquad \qquad \qquad \begin{small} \begin{bmatrix}
 + & - & + \\
 + & +  & - \\
 -  & +  & +
\end{bmatrix} \end{small} \qquad \qquad
\begin{footnotesize} \hbox{Missing nodes $\textcolor{blue}{1}\textcolor{red}{2},\textcolor{blue}{2}\textcolor{red}{3},\textcolor{blue}{3}\textcolor{red}{1}$ from Figure \ref{pic3x3} are negative.} \end{footnotesize}
$$
Our punchline is that this recovers
the split amplitude \eqref{splitk2} for $\mathcal{M}_{0,6}$.
Indeed,  (\ref{eq:tobl1}) is equal~to
\begin{equation}
\label{eq:tobl2}
\frac{u_{\textcolor{blue}{1}\textcolor{red}{1}}+u_{\textcolor{blue}{1}\textcolor{red}{3}}}{ u_{\textcolor{blue}{1}\textcolor{red}{1}} u_{\textcolor{blue}{1}\textcolor{red}{3}}}\, \,\cdot \, \,
\frac{u_{\textcolor{blue}{2}\textcolor{red}{1}}+u_{\textcolor{blue}{2}\textcolor{red}{2}}}{ u_{\textcolor{blue}{2}\textcolor{red}{1}} u_{\textcolor{blue}{2}\textcolor{red}{2}}} \,\,\cdot \, \,
\frac{u_{\textcolor{blue}{3}\textcolor{red}{2}}+u_{\textcolor{blue}{3}\textcolor{red}{3}}}{ u_{\textcolor{blue}{3}\textcolor{red}{2}} u_{\textcolor{blue}{3}\textcolor{red}{3}}}.
\end{equation}
The translation into the coordinates used earlier (for $k=2,n=6)$ is given by
$u_{\textcolor{blue}{1}\textcolor{red}{1}}=s_{12}$,
$u_{\textcolor{blue}{1}\textcolor{red}{2}}=s_{24}$,
$u_{\textcolor{blue}{1}\textcolor{red}{3}}=s_{23}$,
$u_{\textcolor{blue}{2}\textcolor{red}{1}}=s_{234}$,
$u_{\textcolor{blue}{2}\textcolor{red}{2}}=s_{123}$,
$u_{\textcolor{blue}{2}\textcolor{red}{3}}=s_{35}$,
$ u_{\textcolor{blue}{3}\textcolor{red}{1}}=s_{46}$,
 $u_{\textcolor{blue}{3}\textcolor{red}{2}}=s_{61}$, and
 $u_{\textcolor{blue}{3}\textcolor{red}{3}}=s_{56}$.
\end{example}

In this paper we introduced a kinematic subspace on which the CEGM amplitude $m_n^{(k)}$ splits into the product of the simplex amplitudes (Section \ref{sec:simplex}). These are  simplicial degenerations of the standard biadjoint scalar amplitude. This has led to the discovery of a class of hidden zeros of $m_n^{(k)}$. Another noteworthy result of this paper appears in Section~\ref{sec:GSF}: we show how the split CEGM amplitude \eqref{eq:splitamplitude} can be obtained from the tropical determinantal variety ${\cal T}_k^{p\times q}$ of $p\times q$ matrices of rank $k$. We focus on the case of maximal minors, when $p=k+1$ and $q=n-k-1$. In this last section we started the exploration of determinantal varieties for matrices of rank $k=2$, with $p=3$ and $q=n-3$ being the case of maximal minors. This gives rise to the splitting of the $n$-point biadjoint scalar amplitude shown in \eqref{splitk2}. This connects to the space of bicolored trees, which in Lam's theory \cite{Lam:2024jly} corresponds to the Bergman fan of 
a bipartite graphic matroid, and the split amplitude is the Laplace transform over the fan. 

It would be very interesting to extend the study of scattering amplitudes from
determinantal varieties in the context of Lam's theory to more general cases. For example, one can start with $4\times q$ matrices of rank $2$. The motivation behind this is that it connects to the space of metric trees with four leaves and $q$ marked points, so one would hope to obtain rational functions that might be equivalent to summing over a certain class of Feynman diagrams with $4+q$ particles. More generally, one would expect the variety ${\cal T}_k^{p\times q}$ to extend to the CEGM theory, as this relates to the moduli space of tropical $(k-1)$-hyperplanes in the tropical projective space $\mathbb{TP}^{p-1}$ with $q$ marked points. In future work we will continue to build on this new exciting line of ideas.

\section*{Acknowledgements}
We are grateful to Nick Early for pointers and conversations.
BGU thanks the Max Planck Institute for Mathematics in the Sciences and the Institute for Advanced Study for their hospitality while this work was completed. BGU is an STFC Research Fellow supported by the STFC consolidated grant ST/X000583/1. BSt is supported by the European Union (ERC-Synergy Grant UNIVERSE+, 101118787).
\begin{footnotesize}
Views and opinions expressed are however those~of the authors only and do not
necessarily reflect those of the European Union or the European Research Council
Executive Agency. Neither the European Union nor the granting authority can
be held responsible for them.
\end{footnotesize}

\bibliographystyle{jhep}
\bibliography{references}

\end{document}